\documentclass[
    ,final            
  ]
  {aipproc}

\layoutstyle{6x9}

\def\aap{\mbox{{A\&A}}}
\def\apj{\mbox{{ApJ}}}

\def\mnras{\mbox{{MNRAS}}}

\newcommand{\Msun}{\mbox{$\mathrm{M}_{\odot}$~}}

\begin{document}

\title{Critically rotating stars in binaries \\ - an unsolved problem -}


\author{S.~E. de Mink}{
  address={Astronomical Institute Utrecht,
              Princetonplein 5, NL-3584 CC Utrecht, The Netherlands,
              S.E.deMink@astro.uu.nl}
}

\author{O.~R. Pols}{}

\author{E. Glebbeek}{}

\begin{abstract}
In close binaries mass and angular momentum can be transferred from
one star to the other during Roche-lobe overflow. The efficiency of
this process is not well understood and constitutes one of the
largest uncertainties in binary evolution.

One of the problems lies in the transfer of angular momentum, which
will spin up the accreting star.  In very tight
systems tidal friction can prevent reaching critical rotation, by
locking the spin period to the orbital period. 
Accreting stars in systems with orbital periods larger than a few days
reach critical rotation after accreting only a fraction of their mass,
unless there is an effective mechanism to get rid of angular
momentum. In low mass stars magnetic field might help. In more massive
stars angular momentum loss will be accompanied by strong mass loss.
This would imply that most interacting binaries with initial orbital
periods larger than a few days evolve very non-conservatively.

In this contribution we wish to draw attention to the unsolved
problems related to mass and angular momentum transfer in binary
systems.  We do this by presenting the first results of an
implementation of spin up by accretion into the \texttt{TWIN} version
of the Eggleton stellar evolution code.

\end{abstract}

\maketitle


\section{Introduction}
The majority of stars are found in binary systems and a large fraction
of them are so close that the two stars interact during their lifetime
by exchanging mass. This completely alters the evolution of both stars
compared to that of isolated stars.

Model calculations of mass transfer in binaries have been around since
almost 40 years and have succesfully explained the main
characteristics of for example Algol systems, binaries in which the
less massive star is more evolved than the more massive star. In these
first models it was commonly assumed that mass transfer is a
conservative process, i.e. neither mass nor angular momentum is lost from
the system. It has become clear that this picture is not valid, at
least for some systems.

The first approaches to model non-conservative mass transfer assumed
that a fraction $\beta$ of the transferred mass is lost from the
system, where $\beta=0.5$ has been commonly adopted regardless of the physical
mechanism behind the mass loss or the properties of the binary system.

In \cite{demink+ea07} we compared a large grid of detailed binary
models, making different assumptions for $\beta$, to observations of
double lined eclipsing binaries. We found poor agreement when adopting
one constant value for $\beta$. The slightly wider systems in the
observed sample seem to have evolved less conservatively than the
closer systems. We speculated that this might be explained by the fact
that in close systems tidal forces can keep the accreting star in
synchronous rotation with the orbit, while in the wider systems the
accreting star is spun up by the accretion stream \citep{Packet81}. As
it rotates faster it experiences enhanced mass loss \citep{Langer98},
which may explain why the wider systems evolved less conservatively.

The need for a more physical description of non-conservative mass
transfer was already suggested by \cite{Wellstein01, Langer_ea03,
Petrovic_ea05}. They assume that mass loss in the form of stellar
winds is enhanced by the rotation. Mass is accreted untill the star
reaches critical rotation.

We recently implemented a model of spin up by mass transfer in the
\texttt{TWIN} code, a detailed binary evolution code suitable for
calculating large grids of binary models if used on a computer
cluster. In this contribution we present the first results.

\section{Implementation of Spin up in the Evolution Code}

The \texttt{TWIN} code \citep{Eggleton_ea98, Eggleton+Kiseleva02} is a
binary evolution code based on the \texttt{STARS} code
\citep{Eggleton71, Eggleton72, Pols+ea95}. It solves the structure and
composition equations for the two stars in a binary simultaneously
with equations for the orbit assuming rigid rotation.
Non-conservative mass transfer is implemented by assuming that a
constant fraction $\beta$, a free parameter, of the transferred mass
is lost from the system. After showing that observations do not support
a constant fraction \citep{demink+ea07} we implemented a more realistic
model of angular momentum transfer in the TWIN code.

We distinguish between accretion from a disk and accretion by a direct
impact stream. We use an analytic fit by \cite{Ulrich+Burger76} to
calculations of \cite{Lubow+Shu75}, which give the minimum distance
$R_{\rm min}$ between the mass transfer stream and the center of mass
of the accreting star as function of the separation and the mass
ratio.  By comparing the radius of the accreting star $R_{\rm A}$ to
the minimum distance of the stream we determine whether disk
accretion, when $R_{\rm A} < R_{\rm min}$, or direct impact accretiom,
when $R_{\rm A} > R_{\rm min}$, occurs.  In the case of disk accretion
we assume that material is accreted with the specific angular momentum
equivalent to that of a Keplerian orbit with the radius of the
accreting star,
\[
h = \sqrt{G M_{\rm A} R_{\rm A}}.
\] 
In the case of impact we assume the specific angular momentum is
that of the Keplerian disk that would have formed if the accreting
star had been a point mass,which has a radius of $1.7R_{\rm \min}$
\citep[according to][]{Lubow+Shu75}, so that 
\[
h = \sqrt{G M_{\rm A} 1.7 R_{\rm min}}.
\]
For the mass losing star we assume that the material is lost from the
inner Lagrangian point with specific angular momentum 
\[
h = \omega_{\rm D} R_{L_1}^2 ,
\] 
where $\omega_{\rm D}$ is the angular speed of rotation of
the donor star and $R_{L_1}$ is the distance of the center of mass of
the donor star to the inner Lagrangian point.
We implement enhanced mass loss $\dot{M}(\omega)$ for stars rotating
at a fraction $\omega/\omega_{\rm cr}$ of critical rotation as in
\citet{Langer98}: 
\[
\dot{M}(\omega) = \dot{M_0} *(1-\omega/\omega_{\rm cr})^{-0.43},
\]
where $\dot{M_0}$ is the mass loss of a non-rotating star. When the the
rotation rate of the accreting star reaches 99\% of critical rotation
we prevent any more accretion. At this moment our code experiences
convergence problems when the accreting star rotates too fast so we
stop our calculations. We hope to solve this in the near future.

\begin{figure*}    
\includegraphics[ 
        width=\columnwidth]{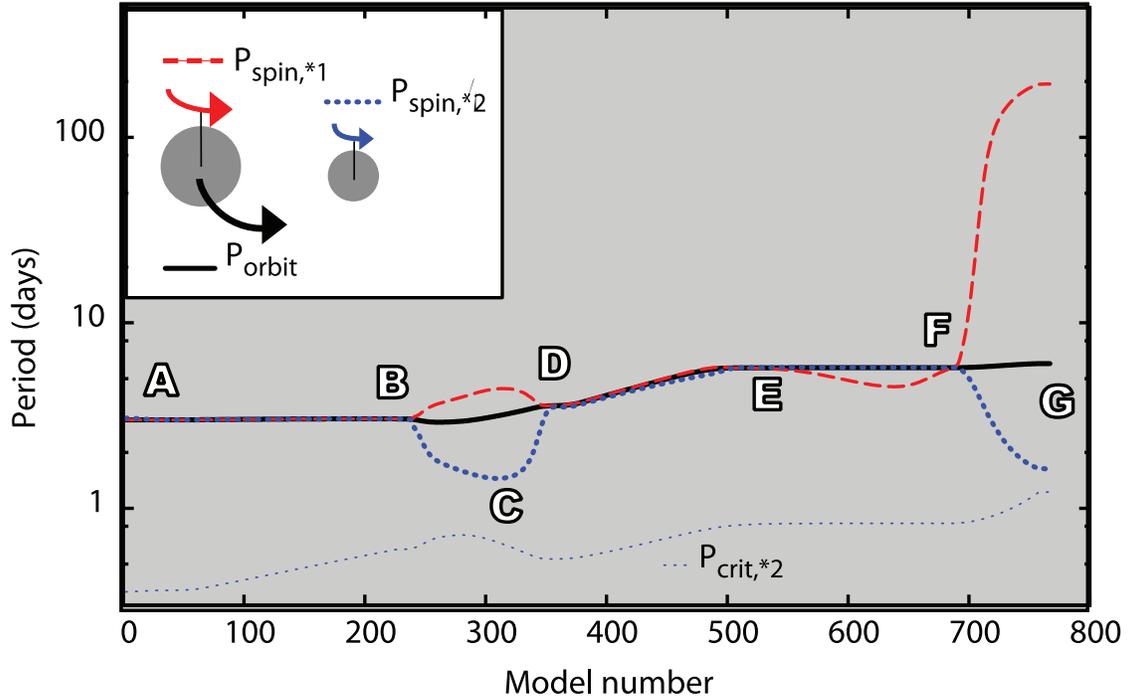}
\caption{ Example of the evolution of the orbital period $P_{\rm
orbit}$ and the spin periods $P_{\rm spin, *1,2}$ of a binary
consisting of a 20 \Msun and a 16 \Msun star. The labels are explained
in the main text.
        \label{graph} }
   \end{figure*}

\section{First Results}
As an example we show the evolution of a massive binary, consisting of
a 20 \Msun and a 16 \Msun star, with an initial orbital period of 3
days, in such a close orbit that the tidal forces can prevent the
accreting star from reaching critical rotation in the first phases of
mass transfer. Figure~\ref{graph} shows the orbital period and the
spin periods of both stars against the number of the computed model,
which is essentially a non-linear time axis stretching rapid phases of
the evolution.

 We start the evolution with the spin periods of both stars
synchronized and aligned with the orbit (A). The primary star expands
as it evolves on the main sequence and fills its Roche lobe (B). It
starts to transfer mass and angular momentum on a thermal time scale
to its companion.  During this rapid phase of mass transfer the
accreting star spins up (blue), while the donor spins down (red)
(C). When the mass ratio is reversed the orbit widens and the mass
transfer rate slows down. Tidal interaction again synchronizes the
rotation of the stars with the orbit (D). Mass transfer continues on a
nuclear time scale. At (E) the primary has burned all its central
Hydrogen and starts to contract at the end of the main sequence until
hydrogen ignites in a shell and the star expands on a thermal
timescale as it crosses the Hertzsprung gap (F). Mass is now being
transferred on a thermal timescale. The accreting star is spun up
close to critical rotation and our simulation ends (G).

\section{Unsolved Problems }

As first shown by \citet{Packet81}, angular momentum transfer is so
efficient that a star can reach critical rotation after accreting only
a small fraction of its own mass. In systems slightly wider than the
example discussed above, this effect will severely limit the amount of
mass that can be accreted by the secondary, and will lead to very
non-conservative binary evolution.  While this is consistent with some
observed binaries, there are several counterexamples indicating that
even in rather wide post-mass transfer binaries, with periods of
$100$~days or more, mass transfer has been fairly conservative,
e.g. $\phi$\,Per \citep{Pols07}.


In order for fairly conservative mass tranfer to be possible in all
but the closest binaries, an effective angular momentum loss mechanism
must operate.  We briefly discuss several such mechanisms below,
noting that how effective most of these mechanisms are is very
uncertain.

\begin{description}

\item[Tidal interaction] tends to keep the spin period of the accreting star
synchronized with the orbit. However, tides are not efficient enough
during rapid mass transfer or in systems wider than a few days, as
shown in the example above.

\item[Rotation-enhanced wind mass loss] and the associated angular
momentum loss in massive binaries with strong intrinsic winds can slow
down the star when mass is lost preferentially in the equatorial plane
or spin up the star when it is lost preferentially at the poles.  In
both cases it leads to highly non-conservative evolution.

\item[Mass shedding] of accreted material from the equator when the 
rotation is close to critical might work for intermediate-mass and/or
low-metallicity binaries. Similar to rotation-enhanced wind mass loss,
it should lead to very non-conservative mass transfer. Since
conservative evolution seems possible even in fairly wide systems, a
mechanism to lose angular momentum without much mass loss must exist.

\item[Delayed accretion]. Perhaps the transferred mass and
angular momentum can be stored temporarily in a circumstellar or
circumbinary disk. In an accretion disk mass can be transported
inwards, while angular momentum is transported outwards and eventually
transferred back to the orbit by tidal interaction with the disk. This
allows the possiblity to accrete the stored mass on a slower timescale
than the rapid mass transfer timescale. If this process is effective,
it can conserve both the mass and the angular momentum of the binary.

\item[Magnetic fields] have the potential to carry angular momentum
away from the star with relatively little mass loss. This process of
magnetic braking works efficiently for low-mass stars with convective
envelope, so that low-mass binaries can lose angular momentum without
losing much mass. However, massive main sequence stars with radiative
envelopes do not generally have strong magnetic fields, with a few
exceptions \citep{Henrichs+ea05}.  This process might be effective if
strong magnetic fields can be generated during the accretion process
itself.

\end{description}
Many open questions remain. Future research should address simple but
proper ways to model angular momentum loss mechanisms, which are not
well understood, and a comparison of such models to observed post-mass
transfer binaries with well-determined parameters.


\begin{theacknowledgments}
 We would like to thank Peter Eggleton for
his stellar evolution code. SdM thanks LKBF for financial support to
attend this conference.
\end{theacknowledgments}
\bibliography{C07_spinup}


\bibliographystyle{mn2e}


\IfFileExists{\jobname.bbl}{}
 {\typeout{}
  \typeout{******************************************}
  \typeout{** Please run "bibtex \jobname" to optain}
  \typeout{** the bibliography and then re-run LaTeX}
  \typeout{** twice to fix the references!}
  \typeout{******************************************}
  \typeout{}
 }

\end{document}